# Binding memory of liquid molecules

Shiyi Qin[1,3], Zhi Yang[1,3], Huimin Liu[1], Xiaoli Wang[1,2], Shangguo Hou[1,]*, Kai Huang[1,]*

[1]Institute of Systems and Physical Biology, Shenzhen Bay Laboratory, Shenzhen, 518132, China
[2]Department of Chemistry, College of Sciences, Northeastern University, Shenyang 110819, China
[3]These authors contribute equally
*Correspondence: huangkai@szbl.ac.cn (K.H.), shangguo.hou@szbl.ac.cn (S.H.)

**Abstract:** Understanding the binding dynamics of liquid molecules is of fundamental importance in physical and life sciences. However, nanoscale fast dynamics pose great challenges for experimental characterization. Conventionally, the binding dynamics have been assumed to be memoryless. Here, we integrate large scale computer simulation, scaling theory, and real-time single particle tracking microscopy with high spatiotemporal precision to unveil a universal memory effect in the binding dynamics of liquid molecules. This binding memory can be quantified by a binding time autocorrelation function, whose power-law decay depends not only on the binding affinity, but also on the topological and materials properties of the surrounding environment. Context-dependent biomolecular binding memory is likely exploited by biological systems to regulate biochemical reactions and biophysical processes. Deciphering this binding memory offers a novel strategy to probe complex biological systems and advanced soft materials.

**Introduction**

The dynamics of molecular binding in liquid environments is of critical importance across a variety of fields, from chemical physics to biophysics and nanotechnology. For instance, the initial viral infection involves binding to the cell membrane and transcription factors must reside at cis-regulatory elements to activate gene expression. On the other hand, engineering molecular and nanoparticle binding is key to optimizing material properties with applications in catalysis[1,2], self-assembly[3,4], stimuli-responsiveness[5,6], and drug delivery[7]. Despite its fundamental importance, our understanding of the binding dynamics is still limited due to the experimental difficulty to capture the fast molecular binding events.

Since a single binding event may not always trigger downstream activities, it is necessary to consider the impact of rebinding dynamics. In many soft matter systems, particularly biological ones, functionality often depends on a sequence of weak binding and rebinding events rather than a single, strong interaction. For instance, recent study revealed that frequent "on-and-off" binding of transcription factors at cis-regulatory element is associated with prolonged transcriptional bursts, suggesting that rebinding is functionally important in gene regulation[8]. Other studies reported that rebinding has profound influences on enzymatic reactions[9], ligand recogniztion[10,11] and drug binding affinity[12]. Nevertheless, aside from these observations, it remains largely unknown how frequent rebinding happens and how it shapes the time dependence of molecular binding probability.

It is widely believed that the binding probability decreases exponentially over time, with the rate of decay characterized by a dissociation rate constant $k_{off}$[13,14]. The underlying assumption is that molecular binding dynamics is stochastic and memoryless. This assumption is deeply rooted and a corner stone in almost every model that concerns molecular binding dynamics. For example, continuum affinities models have been developed to explain non-exponential binding dynamics based on the superposition of multiple single-exponential functions[15–17]. In other cases, an "apparent" $k_{off}$ is introduced to account for observed binding lifetime that is much longer than theoretical prediction[18–20]. These empirical models provide only limited molecular insights, as they cannot distinguish between single strong binding and multiple weak binding/rebinding events. To fully reconcile the discrepancies between theory and experiment, a new paradigm is needed wherein the effect of rebinding is properly incorporated.

Here, we propose that the intricate interplay between the binding kinetics and the molecular dynamics can lead to a binding memory effect. Combining simulation, theory and experiment, we show that binding memory is ubiqituous in liquid and may play important roles in functional soft matter systems. We develop a general theoretical framework to understand this memory effect, highlighting its dependence on various factors. Finally, we demonstrate how to probe soft matter by deciphering the molecular binding dynamics and discuss the biological implications of molecular binding memory.

**The concept of molecular binding memory**

The idea of molecules having binding memory might seem odd, given the long-held assumption that nanoscale particles are subjected to memoryless, random forces, as described by Einstein's renowned theory of Brownian motion. However, it was not until more than half a century later that scientists realized microscopic hydrodynamics could induce a memory effect in the motion of Brownian particles. The overlooked hydrodynamic memory was first discovered in the seminar computer simulation by Alder and Wainwright[21], and later captured by optical tweezer experiments[22,23].

If Brownian motion has memory, what about the binding of molecules and nanoparticles? Can the concept of memory be extended to molecular interactions? To explore this, we first need a mathematical definition of binding memory that can be analyzed theoretically and measured experimentally. A natural choice is the time autocorrelation function, also known as the memory kernel. For example, hydrodynamic memory is quantified by the velocity time autocorrelation function of the Brownian particle. Here, we introduce the binding time autocorrelation function (BAF) to study the molecular binding memory, which can be expressed as:

$$\text{BAF}(\Delta t) = \frac{\langle n(t+\Delta t)n(t)\rangle}{\langle n(t)n(t)\rangle} \quad (1)$$

where $n(t)$ represents the number of bound molecules at time $t$, and $n(t+\Delta t)$ counts the molecules that still reside at their binding sites at moment $t+\Delta t$. The time lag $\Delta t$ is a positive number. The bracket means ensemble averaging, which makes BAF $t$-independent.

Unlike the lifetime distribution function (LDF) that concerns only the duration of a single binding event, the BAF accounts for the contribution of rebinding events to the overall binding probability (Fig. 1). A memoryless BAF would reduce to the LDF, which typically follows an exponential decay. However, molecules with binding memory will exhibit BAFs that decay slower than an exponential function. Besides BAF, bindng memory can also be characterized by other quantities as we will discuss later.

**All-atom molecular dynamics simulations unveil binding memory effect**

Computer simulation has been proven a powerful tool to investigate molecular autocorrelation functions[21,24,25]. However, accurately calculating the BAF is computationally expensive, as it requires continuous tracking of the binding dynamics among a vast ensemble of molecules. Moreover, the periodic boundary condition of the simulation box can lead to artifact of BAF at long-time scale (Fig. S1). We address these problems and accelerate the data processing by about three orders of magnitudes via optimizing the molecular neighboring analysis (see the

Methods for more details). By using this technique and performing large-scale simulations (particle number up to 3.75 million) over more than 1000 times, we obtain high-quality statistics of the BAFs across multiple timescales.

We first conduct all-atom molecular dynamics (MD) simulations of associated liquid rich in hydrogen bonds, such as water and alcohol. The BAFs of hydrogen bonding in systems of different dimensions are analyzed and shown in Fig. 1B. In these log-log plots, all the BAFs are clearly non-exponential with long-time tails following power law functions. The scaling exponents of the power law functions unequivocally align with $-D/2$, where $D$ is the dimension of the system. The dimension-dependent power law scaling is universal, as it does not depend on the type of the associated liquid or the specific definition of its hydrogen bonding (Fig. 1, S2).

It is intriguing that ethanol manifests a stronger binding memory than water, despite having weaker hydrogen bonds. This result highlights the fact that the collective behavior of liquid molecules may have a greater impact on the binding memory than their individual binding affinities. Water, with its extensive hydrogen bonding network and high molecular mobility, allows for rapid rearrangement and exchange of neighbors, leading to a short memory of previous binding partners. While our all-atom simulations are limited to small solvent molecules, the significant increase of binding memory from water to ethanol suggests that the memory effect could be much more pronounced for large solute molecules.

**Langevin dynamics simulations identify factors that control binding memory**

The observed universality of the BAF scaling could be attributed to the conservation laws such as mass conservation (governing diffusion) and momentum conservation (governing hydrodynamics), both of which are implemented in all-atom MD. Using Langevin dynamics (LD) simulations without hydrodynamics, we further show that diffusion alone can preserve the dimension-dependent scaling law of BAF (Fig. S3). The high efficiency of the LD simulation allows us to systematically study the dependence of BAF on a wealth of factors including viscosity (Fig. 2A), crowding (Fig. 2B), molecular weight (Fig. S4), binding strength and range (Fig. 2C, S5), as well as residence cutoff distance (Fig. 2D). All these factors affect the amplitude of the BAF rather than the scaling exponent.

The above computer simulations suggest that the scaling exponent and amplitude of the BAF encode different charateristics of the liquid environment, with the former being sensitive to the dimensionality and the latter sensitive to the material properties of the liquid as well as the binding affinity. While the binding affinity largely determines the LDF, it is only one of the many factors that shape the BAF. The fact that the binding affinity does not affect the scaling exponent of the BAF has an important implication, i.e., mixing different binding affinities does not break the power law. Neither is the scaling exponent of the power law dependent on the residence cutoff distance, which means that changing the imaging resolution will not alter the observed scaling behavior.

The universal scaling of BAF, invariant of binding energy and imaging resolution, is in striking contrast with the exponential LDF, stressing the importance of rebinding events that have been often neglected in biophysical models. By sorting and analyzing individual binding profiles, we show that rebinding heavily contributes to the effective binding time (Fig. 2E). Ignoring rebinding can lead to significant underestimation of the effective binding time, especially in low-dimensional systems such as cell membranes. It is worth noting that the dimension of the system could change across the length scale, with confinement being an example of decreasing dimensionality. The change of dimension due to confinement can lead to a transition in the scaling of BAF (Fig. S6).

**Scaling theory explains the power law of binding probability decay**
We then carry out theoretical analysis to gain deeper understanding of the scaling of the BAF. While the exact BAF is difficult to predict analytically, its scaling exponent can be determined using scaling theory[26], a powerful tool widely used in polymer physics. Our simulation results indicate that the asymptotic behavior of the BAF is dictated by diffusion, and the effect of binding kinetics can thus be absorbed into a time-independent term at long-time scale that only contributes to the amplitude of the power law. This separation allows us to study the scaling exponent in the extreme case of vanishing binding energy. In this case, the trajectory of liquid molecule can be mapped onto the conformation of a polymer. Therefore, analysis of the BAF is converted to solving the contact probability function $P_c(l)$ that describes the likelihood of contact between two loci of the trajectory "polymer" as a function of the contour length between them. Here, the contour length $l$ is equivalent to the time $t$ in the BAF context.

The scaling property is associated with the concept of fractal[27], i.e., self-similarity across scales. Not to loss generality, we assume that the trajectory "polymer" is of dimension $\theta$ (the dimension of the walk), embedded in a $D$-dimensional space. The contact probability function $P_c(l)$ follows a power law

$$P_c(l) = Al^{-s} \qquad (2)$$

where $s$ is also the scaling exponent of BAF that we are interested in. Two loci can be said to be in contact if they are within a contact cutoff distance $b$, or $b/\varepsilon$ in the measure of the scale unit $\varepsilon$. Changing the contact criterion will rescale the amplitude of the contact probability function

$$A(b, \varepsilon) = A_0 (b/\varepsilon)^D \qquad (3)$$

where $A$ is a constant that does not depend on the contact distance. On the other hand, the contour length of a fractal is scale-dependent and can be written as

$$l(\varepsilon) = l_0 \varepsilon^{-\theta} \qquad (4)$$

with $l_0$ being scale-invariant. Subsituting Eq. 3 and Eq. 4 into Eq. 2, we have

$$P_c(l_0, b) = A_0 b^D l_0^{-s} \varepsilon^{\theta s - D} \qquad (5)$$

Since $P_c(l_0, b)$ is scale-invariant, $\theta s - D$ must vanish, leading to

$$s = D/\theta \qquad (6)$$

For normal diffusion we have $\theta = 2$. In a more general situation that considers anomalous diffusion, we have $\theta = 2/\alpha$, where $\alpha$ is the diffusive exponent. Therefore Eq. 6 can be rewritten as

$$s = \frac{\alpha D}{2} \qquad (7)$$

The above theoretical analysis explains the dimension-dependent scaling behaviors observed in our computer simulations (Fig. 1), where both $D$ and $\theta$ are integers.

**The scaling relationship holds in fractal systems**
In principle, both numerator and determinator in Eq. 6 can be non-integer, i.e., fractal dimensions. Fractal space[28–30] and anomalous diffusions[31–33] are often encountered in biological liquid systems. To test the applicability of our theory in complex systems, we have constructed two different liquid environments featuring Hilbert curve (Fig. 3A) and diffusion-limited aggregation (DLA)[34,35] (Fig. 3B), respectively. In the former case, the Hilbert curve is a wall of fixed particles that cannot be penetrated by the mobile molecules whose binding memory is of our interest. In the latter case, the mobile molecules are confined within the DLA. Both fractal structures are embedded in a 2D space. The mobile molecules are adhesive to each other, undergoing dynamic binding and dissociation. We calculate the fractal dimension $D$ of the environments based on mass scaling and analyze the diffusion of the mobile particles to obtain $\theta$ or $\alpha$ (Fig. S7). We then use Eq. 6 to predict the scaling exponent of the BAF. The simulated results, i.e., the ground truth, are shown in Fig. 3C. Our theoretical predictions are in good agreement with our simulation results (Fig. 3D). This agreement is remarkable as it proves that our theory of BAF scaling is not limited to simple topologies or normal diffusion processes, but also applicable to describe the intricate binding dynamics in sophisticated scenarios. Given the complexity of biological systems, the scaling exponents of biomolecular BAFs are expected to cover a continuous spectrum, rather than a few discrete values. Once the scaling exponent is obtained, it can be used to infer the fractal dimension of the environment based on our theory. Such inference offers practical advantages because it is non-invasive and more convenient than directly characterizing the whole environment.

## High-resolution single-particle tracking experiments revealed strong binding memory on artificial and live-cell membranes

Having validated the generality of our theory, we now integrate it with high spatiotemporal precision single-particle tracking (SPT) experiments to characterize complicated liquid environment. Recently, we developed a real-time 3D single particle tracking system with one microsecond temporal resolution and nanometers spatial resolution, termed Single Metal-nanoparticle Active Real-time Tracking with Enhanced Resolution (3D-SMARTER) microscopy[36] (Fig. 4A). The high spatiotemporal resolution of 3D-SMARTER enable it to extract precise biomolecular dynamics in vitro and in live-cell environments.

Here we employed 3D-SMARTER to track the dynamics of gp120-modified silver nanoparticles (gp120-AgNPs) on a supported lipid bilayer (SLB). The gp120 protein, a glycoprotein on the surface of the Human Immunodeficiency Virus (HIV) envelope, plays a critical role in viral infection by binding to the CD4 receptor on T cells, initiating viral entry. By using the depolarized scattering signal of AgNP, we tracked the probe motion, with the 2D trajectory shown in Fig. 4B. Futhermore, the density maps of residence time are extracted in both 2D and 3D with 1 µs resolution (Fig. 4C, D). The high spatiotemporal resolution of 3D-SMARTER allows us to easily distinguish BAF and LDF of the molecule as shown in Fig. 4E, F. Our technique shows that the LDF decays fast and can be fitted to a single-exponential function, whereas the BAF decays much more slowly, exhibiting a power law at long timescales with a scaling exponent of -0.82. The dramatic difference between BAF and LDF revealed by our experiment suggests a high frequency of rebinding events, and highlights the significance of binding memory in real biological systems. In addition, the nanoparticle exhibited sub-diffusive behavior, characterized by an anomalous diffusion exponent of $\alpha = 0.85$. Based on our theory (Eq. 7), we predict that our probe experiences a fractal environment of $D = 1.93$ (Fig. 4F), which is very close to 2, indicating that the lipid membrane is topologically simple.

We then applied 3D-SMARTER to track the residence dynamics of TAT-peptides functionalized silver nanoparticles (TATp-AgNPs) on a live-cell membrane in vivo (Fig. G-I). Unlike the synthetic lipid bilayer, the live-cell membrane is far more complex in its spatial organization, which could lead to nontrivial topological effect. In this system, the LDF is no longer a single-exponential function, but can be better fitted by a bi-exponential function (Fig. 4J), implying the presence of both specific and non-specific interactions with the membrane. The BAF still features a power-law decay, but with a lower scaling exponent of -0.52 (Fig. 4K), indicative of a stronger binding memory on live-cell membranes. Our theoretical model predicted a fractal dimension of $D = 1.22$ for the live-cell membrane, much lower than 2, suggesting a complex topology experienced by the probe. A number of factors could contribute to this topological effect, including heterogeneous distribution of membrane proteins, surface roughness, and even phase separation of the membrane. The difference between live-cell membrane and SLB uncovered by our SPT experiment suggests that lower environmental dimension could be exploited by biological systems to increase molecular binding memory.

**Molecular binding memory beyond single molecule**

So far, we have been focused on molecular binding memory at the single-molecule level, characterized by the BAF. Now we explore the collective memory effect of molecular binding, which may first manifest itself in the fluctuations of equilibrium systems. For an ensemble of molecules, the total number of bound molecules, $N$, fluctuates around its average value, $\overline{N}$, over time (Fig. 5A). The time autocorrelation of this molecular binding fluctuation, $C_f(\Delta t) = \langle (N(t) - \overline{N})(N(0) - \overline{N}) \rangle / \overline{N}^2$, is characterized by computer simulation and a power law is uncovered (Fig. 5B). This result demonstrates that molecular binding fluctuation is not simply random (white noise) but a non-Markovian process with memory effect.

For a population of molecules, the averaged number of intra-population pairing pairs $\langle N_{\text{pair}}(t) \rangle$ decays as a function of time. Ignoring binding memory, this quantity is expected to decay exponentially. Nevertheless, our computer simulations show that, like single-molecule BAF, the population-level pairing number exhibits dimension-dependent power laws (Fig. 5C). Therefore, molecular binding memory is not limited to single molecule or time correlation.

Furthermore, binding memory could manifest in systems out of equilibrium. To study this, we conducted non-equilibrium simulation in which the binding strength is weakened by an external stimuli. Such perturbation leads to a reduction of molecular pairing to a lower level. By tracking the time-dependence of this process in computer simulation, we reveal that the change of molecular pairing number as a function of time follows a power law (Fig. 5D), indicative of a memory effect during the change of system state.

**Biological implications of molecular binding memory**

Biological systems consist of a diversity of biomolecules with specific interactions. Therefore, cognate binding partners for a particular biomolecule is often sparse in space, allowing binding memory to emerge. Moreover, given the high viscosity, crowding level and molecular weight in biological systems, the biomolecular diffusion is often slow, which magnifies the binding memory effect. It is worth noting that many proteins have a great portion of intrinsically disordered regions (IDRs) that increase their hydrodynamic radii. According to the Stokes-Einstein equation, these IDRs further retard the protein diffusion, thereby strengthening protein binding memory.

Binding memory is likely at play at intra-molecular level within IDRs. From a polymer physics point of view, many IDRs can be modeled as associative polymers composed of stickers and spacers[37–39]. The reversible binding of stickers allows dynamic changes of IDR conformations over time. Different sticker-spacer sequences could lead to distinct binding memory effects within the IDRs. To gain insights into this problem, we designed and studied three distinct sticker patterns: uniform, random, and patched, all with the same number of stickers (Fig. 6A). For each sequence, we analyzed the fluctuation of total sticker binding number as a function of time. As shown in Fig. 6A, the time autocorrelation functions of this binding fluctuation exhibit power laws whose scaling exponents depend on the specific sticker pattern. Interestingly, the two well mixed patterns-uniform and random-can be easily distinguished by their scaling exponents, with the former showing a slower decay. Clustering stickers into patches further slows down the dynamics. These results suggest the presence of a memory effect during the conformational changes of IDRs, encoded by the sequence-specific arrangement of stickers.

Many biochemical reactions and assembly of protein complexes occur on biological interfaces, such as membranes[40,41] or surfaces of membraneless compartments[42–44]. Compared to the 3D bulk environment, such interfaces have reduced dimensionalities, sometimes even lower than two dimensions (Fig. 4). Such low dimensionality, combined with the slow interfacial diffusion, could yield strong binding memory effect at biological interfaces. As a proof of concept, we constructed a liquid droplet in our simulation and analyzed the binding memory of molecules trapped to the droplet surface. Our calculation shows that the BAF of interfacial molecules decays much slower than that of the bulk molecules, exhibiting a power law whose scaling exponent is 0.59 (Fig. 6B). Note that the binding memory is stronger than that on a flat interface, indicative of the existence of a curvature effect. The BAF decay flats out at longer timescale due to the limited surface area of the droplet. These findings suggest that the surfaces of biological condensates could play a role in regulating molecular binding memory and downstream biological processes. The sensitivity of binding memory to phase separation also enables the development of sensors for biological condensates using advanced imaging techniques.

**Conclusions and outlooks**

In summary, by integrating large-scale computer simulation, scaling theory, and high-resolution SPT experiment, we have systematically investigated the molecular binding dynamics and unambiguously demonstrated the existence of binding memory in liquid molecules. The binding probability does not decay exponentially over time but follows a power law at long-time scale. The scaling exponent and the amplitude of the power law are found to be sensitive to the topological and material properties of the system, respectively. A universal scaling law has been uncovered to relate the scaling of binding probability to the environmental dimension and the diffusion exponent of the molecule, which enables us to characterize complex liquid environment based on SPT technique. The molecular binding memory have important biological implications as demonstrated by our coarse grained simulations of IDRs and phase separating systems. Unlike chemically fixed binding strength, binding memory is highly tunable and responsive to environmental changes. Such flexibility could have been leveraged by biological systems to regulate critical processes. Consideration of binding memory effect will be important in future biophysical studies and soft materials design.

**Methods**
**Computer simulation and analyses of binding dynamics**

Large-scale and high-throughput simulations allow us to achieve good statistics of binding dynamics. However, it is computationally expensive to extract the information of binding memory from the big data generated by substantial simulations. Employing a hash table for the neighbor list along with OpenMP parallelization[45], we boost the speed of binding analysis by nearly three orders of magnitudes. Amending the artifact of periodic boundary condition (PBC) will inevitably lead to loss of samples over time. To resolve this conundrum, we introduce image binding sites to ensure the correct asymptotic behavior and high statistical quality of the binding memory.

We use all-atom molecular dynamics simulations conducted in GROMACS[46] to study the binding memory of water and ethanol. For one-dimensional confinement, molecules are restricted within a single-walled carbon nanotube of 2.1 nm in diameter, with PBC applied to the axial direction. In the case of two-dimensional confinement, we use two parallel graphene sheets separated by 2.0 nm, applying PBC to the directions parallel to the graphene surfaces. The bulk system is simulated in a cubic box of 16 nm with PBC in all directions.

Langevin dynamics (LD) using LAMMPS[47] is carried out to study the binding memory of coarse-grained proteins in larger systems. The coarse-grained proteins consist of 20 beads interconnected by harmonic springs. The non-bonded isotropic interactions between all monomers are modeled using the standard 12-6 Lennard-Jones potential. In the study of binding memory at liquid droplet surface, time-dependent clustering analysis is performed to select interfacial molecules.

**Set up of 3D-SMARTER**

In 3D-SMARTER, a pair of electro-optic deflectors (Model 310A, Conoptics Inc) and a tunable acoustic gradient lens (TAGLENS-T1, Mitutoyo) are used to generate addressable laser scanning pattern in 3D space after the objective lens. A single-photon avalanche photodiode (APD, SPCM-AQRH-15, Excelitas Technologies) is used to collect the photon signals and record the photon arrival time. The target particle position in the illumination volume can be calculated in real time on a field-programmable gate array board (PCIe-7858, National Instruments Corp) with the photon arrival time information. When the target particle is off the center of illumination volume, a feedback control signal will be sent to the piezoelectric stage (Nano-PDQ275, Nano-OPQ65, Mad City Lab) to move the target particle to the center position. The one microsecond temporal resolution 3D trajectory can be reconstructed with offline recursive Bayesian analysis with the photon arrival time information and laser focus position information.

**3D-tracking gp120-AgNPs on supported lipid bilayer (SLB) with CD4 protein**

Gp120 protein (HY-P70907, MedChemExpress) was mixed with SH-PEG-NHS at a ratio of 1:20 and incubated overnight at 4℃. After purified with a desalting column (89882, Thermo Fisher Scientific), AgNPs (100 nm, AGCB100-1M) were then added to the solution, incubated at room temperature for 2h to obtain the gp120-AgNPs conjugate. Lipid vesicles were prepared by the way of liquid nitrogen freeze-thaw cycles. In brief, The DOPC and DHPC mixture with a molar ratio of 2.5 and 2.5% DGS-NTA (Ni) were repeatedly freeze-thawed in liquid nitrogen and 60℃ water bath for 5-7 times to obtain vesicles for use. For the preparation of SLB, the glass slide was first sonicated with 50% ethanol and deionized water for 15min each, dried with N2, and treated with plasma cleaner. The 30-fold diluted vesicles were immediately added and incubated at 60°C for 20min to form SLB, followed by a wash with Hepes buffer. CD4 protein solution was added to SLB and incubated at room temperature for 1h. After washing with Hepes buffer, gp120-AgNPs were added to it for real-time tracking using 3D-SMARTER.

**3D-tracking of TATp-AgNPs on live-cell membrane**

Streptavidin was first labeled with NHS-PEG-3K-Thiol at a 1:6 ratio, followed by purification. The labeled streptavidin was then mixed with 100 nm AgNPs (AGCB100-1M, NanoComposix) to functionalize the nanoparticles. Subsequently, biotin-TAT peptides (AS-61209) were added to the streptavidin-AgNP complex, and the solution was purified by centrifugation (1600g, 20 min). For the 3D tracking experiments, HeLa cells were first cultured on coverslips overnight. After incubation, the cells were washed three times with cell culture medium and transferred to live-cell imaging solution. TATp-AgNPs were then added to the solution for real-time tracking of their movement on the cell membrane.

**Acknowledgements:** This work was funded by grants from National Natural Science Foundation of China (grant no. 22203055 to K.H., grant no. 22204106 to S.H., grant no. 22403068 to S.Q.), the Major Program of Shenzhen Bay Laboratory (grant no. S241101001 to K.H.) and the Shenzhen Bay Laboratory Open Fund Project (grant no. SZBL2021080601013 to K.H.), the Shenzhen Medical Research Fund (grant no. B2301003 to S.H.). We acknowledge computational support from the Shenzhen Bay Laboratory High Performance Computing and Informatics Core.

**Author contributions:** K.H. and S.H. conceived, guided and supervised the project; S.Q. and Z.Y. designed and performed computer simulations; H.L. and X.W. designed and performed SPT experiments; S.Q. analyzed experimental data; K.H. wrote the manuscript. All authors discussed the results and reviewed the manuscript. Competing interests: The authors declare no competing interests.

**References**
(1) Narlikar, G. J.; Gopalakrishnan, V.; McConnell, T. S.; Usman, N.; Herschlag, D. Use of Binding Energy by an RNA Enzyme for Catalysis by Positioning and Substrate Destabilization. *Proc. Natl. Acad. Sci. U.S.A.* **1995**, *92* (9), 3668–3672. https://doi.org/10.1073/pnas.92.9.3668.
(2) Schwans, J. P.; Kraut, D. A.; Herschlag, D. Determining the Catalytic Role of Remote Substrate Binding Interactions in Ketosteroid Isomerase. *Proc. Natl. Acad. Sci. U.S.A.* **2009**, *106* (34), 14271–14275. https://doi.org/10.1073/pnas.0901032106.
(3) Cawood, E. E.; Guthertz, N.; Ebo, J. S.; Karamanos, T. K.; Radford, S. E.; Wilson, A. J. Modulation of Amyloidogenic Protein Self-Assembly Using Tethered Small Molecules. *J. Am. Chem. Soc.* **2020**, *142* (49), 20845–20854. https://doi.org/10.1021/jacs.0c10629.
(4) Wei, X.; Chen, C.; Popov, A. V.; Bathe, M.; Hernandez, R. Binding Site Programmable Self-Assembly of 3D Hierarchical DNA Origami Nanostructures. *J. Phys. Chem. A* **2024**, *128* (25), 4999–5008. https://doi.org/10.1021/acs.jpca.4c02603.
(5) Li, Y.; Yang, G.; Gerstweiler, L.; Thang, S. H.; Zhao, C. Design of Stimuli-Responsive Peptides and Proteins. *Adv Funct Materials* **2023**, *33* (7), 2210387. https://doi.org/10.1002/adfm.202210387.
(6) Effiong, U. M.; Khairandish, H.; Ramirez-Velez, I.; Wang, Y.; Belardi, B. Turn-on Protein Switches for Controlling Actin Binding in Cells. *Nat Commun* **2024**, *15* (1), 5840. https://doi.org/10.1038/s41467-024-49934-2.
(7) Zhang, Z.; Guan, J.; Jiang, Z.; Yang, Y.; Liu, J.; Hua, W.; Mao, Y.; Li, C.; Lu, W.; Qian, J.; Zhan, C. Brain-Targeted Drug Delivery by Manipulating Protein Corona Functions. *Nat Commun* **2019**, *10* (1), 3561. https://doi.org/10.1038/s41467-019-11593-z.
(8) Pomp, W.; Meeussen, J. V. W.; Lenstra, T. L. Transcription Factor Exchange Enables Prolonged Transcriptional Bursts. *Molecular Cell* **2024**, *84* (6), 1036-1048.e9. https://doi.org/10.1016/j.molcel.2024.01.020.
(9) Aoki, K.; Yamada, M.; Kunida, K.; Yasuda, S.; Matsuda, M. Processive Phosphorylation of ERK MAP Kinase in Mammalian Cells. *Proc. Natl. Acad. Sci. U.S.A.* **2011**, *108* (31), 12675–12680. https://doi.org/10.1073/pnas.1104030108.
(10) Arranz-Plaza, E.; Tracy, A. S.; Siriwardena, A.; Pierce, J. M.; Boons, G.-J. High-Avidity, Low-Affinity Multivalent Interactions and the Block to Polyspermy in *Xenopus l Aevis*. *J. Am. Chem. Soc.* **2002**, *124* (44), 13035–13046. https://doi.org/10.1021/ja020536f.

(11) Oh, D.; Ogiue-Ikeda, M.; Jadwin, J. A.; Machida, K.; Mayer, B. J.; Yu, J. Fast Rebinding Increases Dwell Time of Src Homology 2 (SH2)-Containing Proteins near the Plasma Membrane. *Proc. Natl. Acad. Sci. U.S.A.* **2012**, *109* (35), 14024–14029. https://doi.org/10.1073/pnas.1203397109.
(12) Vauquelin, G.; Charlton, S. J. Long-lasting Target Binding and Rebinding as Mechanisms to Prolong *in Vivo* Drug Action. *British J Pharmacology* **2010**, *161* (3), 488–508. https://doi.org/10.1111/j.1476-5381.2010.00936.x.
(13) Taylor, E. W. Kinetic Studies on the Association and Dissociation of Myosin Subfragment 1 and Actin. *Journal of Biological Chemistry* **1991**, *266* (1), 294–302. https://doi.org/10.1016/S0021-9258(18)52434-0.
(14) Kiel, C.; Filchtinski, D.; Spoerner, M.; Schreiber, G.; Kalbitzer, H. R.; Herrmann, C. Improved Binding of Raf to Ras·GDP Is Correlated with Biological Activity. *Journal of Biological Chemistry* **2009**, *284* (46), 31893–31902. https://doi.org/10.1074/jbc.M109.031153.
(15) Skaug, M. J.; Mabry, J.; Schwartz, D. K. Intermittent Molecular Hopping at the Solid-Liquid Interface. *Phys. Rev. Lett.* **2013**, *110* (25), 256101. https://doi.org/10.1103/PhysRevLett.110.256101.
(16) Skaug, M. J.; Mabry, J. N.; Schwartz, D. K. Single-Molecule Tracking of Polymer Surface Diffusion. *J. Am. Chem. Soc.* **2014**, *136* (4), 1327–1332. https://doi.org/10.1021/ja407396v.
(17) Mazzocca, M.; Colombo, E.; Callegari, A.; Mazza, D. Transcription Factor Binding Kinetics and Transcriptional Bursting: What Do We Really Know? *Current Opinion in Structural Biology* **2021**, *71*, 239–248. https://doi.org/10.1016/j.sbi.2021.08.002.
(18) Biancaniello, P. L.; Kim, A. J.; Crocker, J. C. Long-Time Stretched Exponential Kinetics in Single DNA Duplex Dissociation. *Biophysical Journal* **2008**, *94* (3), 891–896. https://doi.org/10.1529/biophysj.107.108449.
(19) Chen, J.; Zhang, Z.; Li, L.; Chen, B.-C.; Revyakin, A.; Hajj, B.; Legant, W.; Dahan, M.; Lionnet, T.; Betzig, E.; Tjian, R.; Liu, Z. Single-Molecule Dynamics of Enhanceosome Assembly in Embryonic Stem Cells. *Cell* **2014**, *156* (6), 1274–1285. https://doi.org/10.1016/j.cell.2014.01.062.
(20) Asadollahi, K.; Rajput, S.; De Zhang, L. A.; Ang, C.-S.; Nie, S.; Williamson, N. A.; Griffin, M. D. W.; Bathgate, R. A. D.; Scott, D. J.; Weikl, T. R.; Jameson, G. N. L.; Gooley, P. R. Unravelling the Mechanism of Neurotensin Recognition by Neurotensin Receptor 1. *Nat Commun* **2023**, *14* (1), 8155. https://doi.org/10.1038/s41467-023-44010-7.
(21) Alder, B. J.; Wainwright, T. E. Decay of the Velocity Autocorrelation Function. *Phys. Rev. A* **1970**, *1* (1), 18–21. https://doi.org/10.1103/PhysRevA.1.18.
(22) Li, T.; Kheifets, S.; Medellin, D.; Raizen, M. G. Measurement of the Instantaneous Velocity of a Brownian Particle. *Science* **2010**, *328* (5986), 1673–1675. https://doi.org/10.1126/science.1189403.
(23) Kheifets, S.; Simha, A.; Melin, K.; Li, T.; Raizen, M. G. Observation of Brownian Motion in Liquids at Short Times: Instantaneous Velocity and Memory Loss. *Science* **2014**, *343* (6178), 1493–1496. https://doi.org/10.1126/science.1248091.
(24) Luzar, A.; Chandler, D. Hydrogen-Bond Kinetics in Liquid Water. *Nature* **1996**, *379* (6560), 55–57. https://doi.org/10.1038/379055a0.
(25) Huang, K.; Szlufarska, I. Effect of Interfaces on the Nearby Brownian Motion. *Nature Communications* **2015**, *6* (1). https://doi.org/10.1038/ncomms9558.
(26) Gennes, P. G. de. *Scaling Concepts in Polymer Physics*; Cornell University Press: Ithaca, N.Y, 1979.
(27) Mandelbrot, B. B. *The Fractal Geometry of Nature*; W.H. Freeman: San Francisco, 1982.
(28) Récamier, V.; Izeddin, I.; Bosanac, L.; Dahan, M.; Proux, F.; Darzacq, X. Single Cell Correlation Fractal Dimension of Chromatin: A Framework to Interpret 3D Single Molecule Super-Resolution. *Nucleus* **2014**, *5* (1), 75–84. https://doi.org/10.4161/nucl.28227.
(29) Bancaud, A.; Huet, S.; Daigle, N.; Mozziconacci, J.; Beaudouin, J.; Ellenberg, J. Molecular Crowding Affects Diffusion and Binding of Nuclear Proteins in Heterochromatin and Reveals the Fractal Organization of Chromatin. *EMBO J* **2009**, *28* (24), 3785–3798. https://doi.org/10.1038/emboj.2009.340.

(30) Mazzocca, M.; Fillot, T.; Loffreda, A.; Gnani, D.; Mazza, D. The Needle and the Haystack: Single Molecule Tracking to Probe the Transcription Factor Search in Eukaryotes. *Biochemical Society Transactions* **2021**, *49* (3), 1121–1132. https://doi.org/10.1042/BST20200709.
(31) Kaur, G.; Costa, M. W.; Nefzger, C. M.; Silva, J.; Fierro-González, J. C.; Polo, J. M.; Bell, T. D. M.; Plachta, N. Probing Transcription Factor Diffusion Dynamics in the Living Mammalian Embryo with Photoactivatable Fluorescence Correlation Spectroscopy. *Nat Commun* **2013**, *4* (1), 1637. https://doi.org/10.1038/ncomms2657.
(32) Hansen, A. S.; Amitai, A.; Cattoglio, C.; Tjian, R.; Darzacq, X. Guided Nuclear Exploration Increases CTCF Target Search Efficiency. *Nat Chem Biol* **2020**, *16* (3), 257–266. https://doi.org/10.1038/s41589-019-0422-3.
(33) Höfling, F.; Franosch, T. Anomalous Transport in the Crowded World of Biological Cells. *Rep. Prog. Phys.* **2013**, *76* (4), 046602. https://doi.org/10.1088/0034-4885/76/4/046602.
(34) Liu, D.; Zhou, W.; Song, X.; Qiu, Z. Fractal Simulation of Flocculation Processes Using a Diffusion-Limited Aggregation Model. *Fractal Fract* **2017**, *1* (1), 12. https://doi.org/10.3390/fractalfract1010012.
(35) Tenti, J. M.; Hernández Guiance, S. N.; Irurzun, I. M. Fractal Dimension of Diffusion-Limited Aggregation Clusters Grown on Spherical Surfaces. *Phys. Rev. E* **2021**, *103* (1), 012138. https://doi.org/10.1103/PhysRevE.103.012138.
(36) Hou, S.; Zhang, C.; Niver, A.; Welsher, K. Mapping Nanoscale Forces and Potentials in Live Cells with Microsecond 3D Single-Particle Tracking. June 29, 2022. https://doi.org/10.1101/2022.06.27.497788.
(37) Choi, J.-M.; Holehouse, A. S.; Pappu, R. V. Physical Principles Underlying the Complex Biology of Intracellular Phase Transitions. *Annu. Rev. Biophys.* **2020**, *49* (1), 107–133. https://doi.org/10.1146/annurev-biophys-121219-081629.
(38) Martin, E. W.; Holehouse, A. S.; Peran, I.; Farag, M.; Incicco, J. J.; Bremer, A.; Grace, C. R.; Soranno, A.; Pappu, R. V.; Mittag, T. Valence and Patterning of Aromatic Residues Determine the Phase Behavior of Prion-like Domains. *Science* **2020**, *367* (6478), 694–699. https://doi.org/10.1126/science.aaw8653.
(39) Farag, M.; Cohen, S. R.; Borcherds, W. M.; Bremer, A.; Mittag, T.; Pappu, R. V. Condensates Formed by Prion-like Low-Complexity Domains Have Small-World Network Structures and Interfaces Defined by Expanded Conformations. *Nat Commun* **2022**, *13* (1), 7722. https://doi.org/10.1038/s41467-022-35370-7.
(40) Leonard, T. A.; Loose, M.; Martens, S. The Membrane Surface as a Platform That Organizes Cellular and Biochemical Processes. *Developmental Cell* **2023**, *58* (15), 1315–1332. https://doi.org/10.1016/j.devcel.2023.06.001.
(41) Bondar, A.-N.; Lemieux, M. J. Reactions at Biomembrane Interfaces. *Chem. Rev.* **2019**, *119* (9), 6162–6183. https://doi.org/10.1021/acs.chemrev.8b00596.
(42) Fallah-Araghi, A.; Meguellati, K.; Baret, J.-C.; Harrak, A. E.; Mangeat, T.; Karplus, M.; Ladame, S.; Marques, C. M.; Griffiths, A. D. Enhanced Chemical Synthesis at Soft Interfaces: A Universal Reaction-Adsorption Mechanism in Microcompartments. *Phys. Rev. Lett.* **2014**, *112* (2), 028301. https://doi.org/10.1103/PhysRevLett.112.028301.
(43) Chao, Y.; Shum, H. C. Emerging Aqueous Two-Phase Systems: From Fundamentals of Interfaces to Biomedical Applications. *Chem. Soc. Rev.* **2020**, *49* (1), 114–142. https://doi.org/10.1039/C9CS00466A.
(44) Dai, Y.; Chamberlayne, C. F.; Messina, M. S.; Chang, C. J.; Zare, R. N.; You, L.; Chilkoti, A. Interface of Biomolecular Condensates Modulates Redox Reactions. *Chem* **2023**, *9* (6), 1594–1609. https://doi.org/10.1016/j.chempr.2023.04.001.
(45) *Parallel Programming in OpenMP*; Chandra, R., Ed.; Morgan Kaufmann Publishers: San Francisco, CA, 2001.

(46)	Van Der Spoel, D.; Lindahl, E.; Hess, B.; Groenhof, G.; Mark, A. E.; Berendsen, H. J. C. GROMACS: Fast, Flexible, and Free. *J Comput Chem* **2005**, *26* (16), 1701–1718. https://doi.org/10.1002/jcc.20291.
(47)	Thompson, A. P.; Aktulga, H. M.; Berger, R.; Bolintineanu, D. S.; Brown, W. M.; Crozier, P. S.; In 'T Veld, P. J.; Kohlmeyer, A.; Moore, S. G.; Nguyen, T. D.; Shan, R.; Stevens, M. J.; Tranchida, J.; Trott, C.; Plimpton, S. J. LAMMPS - a Flexible Simulation Tool for Particle-Based Materials Modeling at the Atomic, Meso, and Continuum Scales. *Computer Physics Communications* **2022**, *271*, 108171. https://doi.org/10.1016/j.cpc.2021.108171.

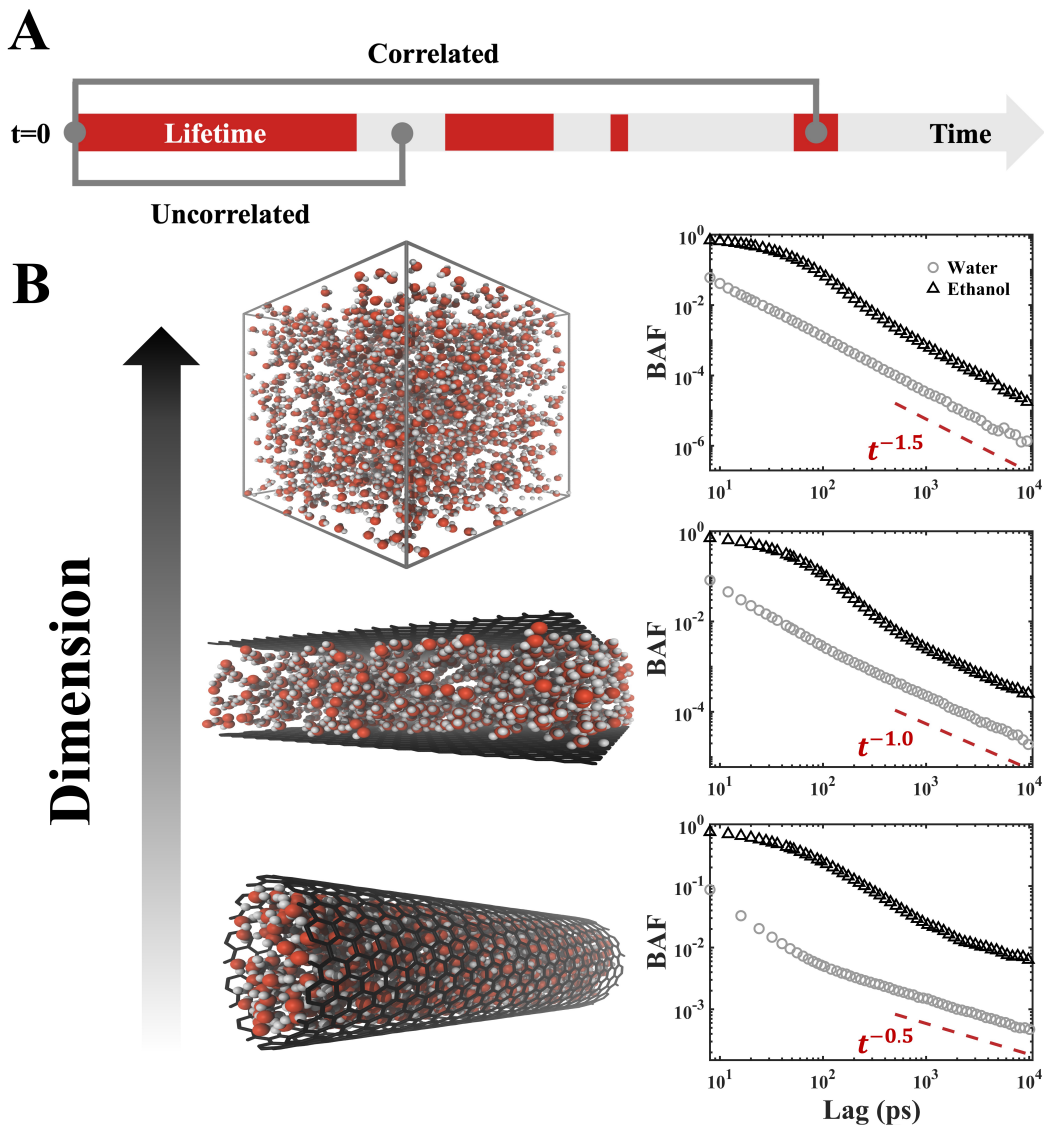

**Fig 1. All-atom simulations revealed topology-dependent molecular binding memory.** (**A**) Schematic definitions of lifetime distribution function and binding time autocorrelation function. Bound state is marked in red and unbound in grey. (**B**) Hydrogen bonding BAFs of water and ethanol in systems of different dimensions (1D to 3D from the bottom to the top). All systems exhibit long-time tails that can be fitted to a universal power law of $t^{-D/2}$, where $D$ represents the dimensionality of the system.

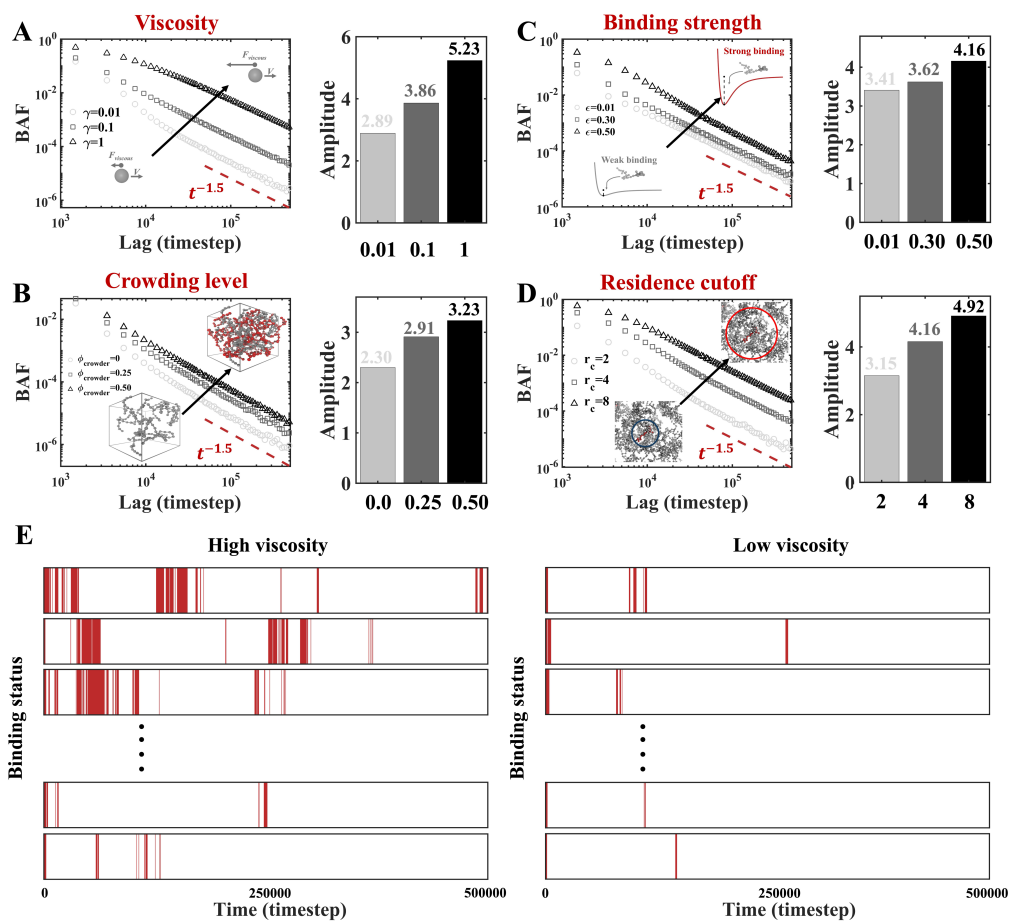

**Fig 2. Effects of non-topological factors on binding memory.** (**A**) The effect of liquid viscosity on molecular BAFs in a three-dimensional system, modeled by changing friction coefficients in Langevin dynamics simulations. Large viscosity enhances molecular binding memory. (**B**) Crowded liquid systems tend to have strong binding memory effect. (**C**) Strengthening binding energy increases binding memory. Note that molecular binding affinity affects the amplitude rather than the scaling exponent of the BAF. (**D**) The amplitude of the BAF power law depends on the choice of binding criterion, i.e., the residence cutoff distance. The power-law amplitudes in bar charts (A-D) are obtained by linear fitting of the BAFs in double-logarithmic plots and taking the intercepts at $\log(t) = 0$, or $t = 1$ in timesteps. (**E**) Rebinding heavily contributes to the effective binding time, demonstrated by comparing liquid systems of different viscosities. The binding dynamics profiles at single-molecule level are sorted by the accumulated binding time within 500,000 simulation steps. The starting points are always in the binding state, which is marked in red.

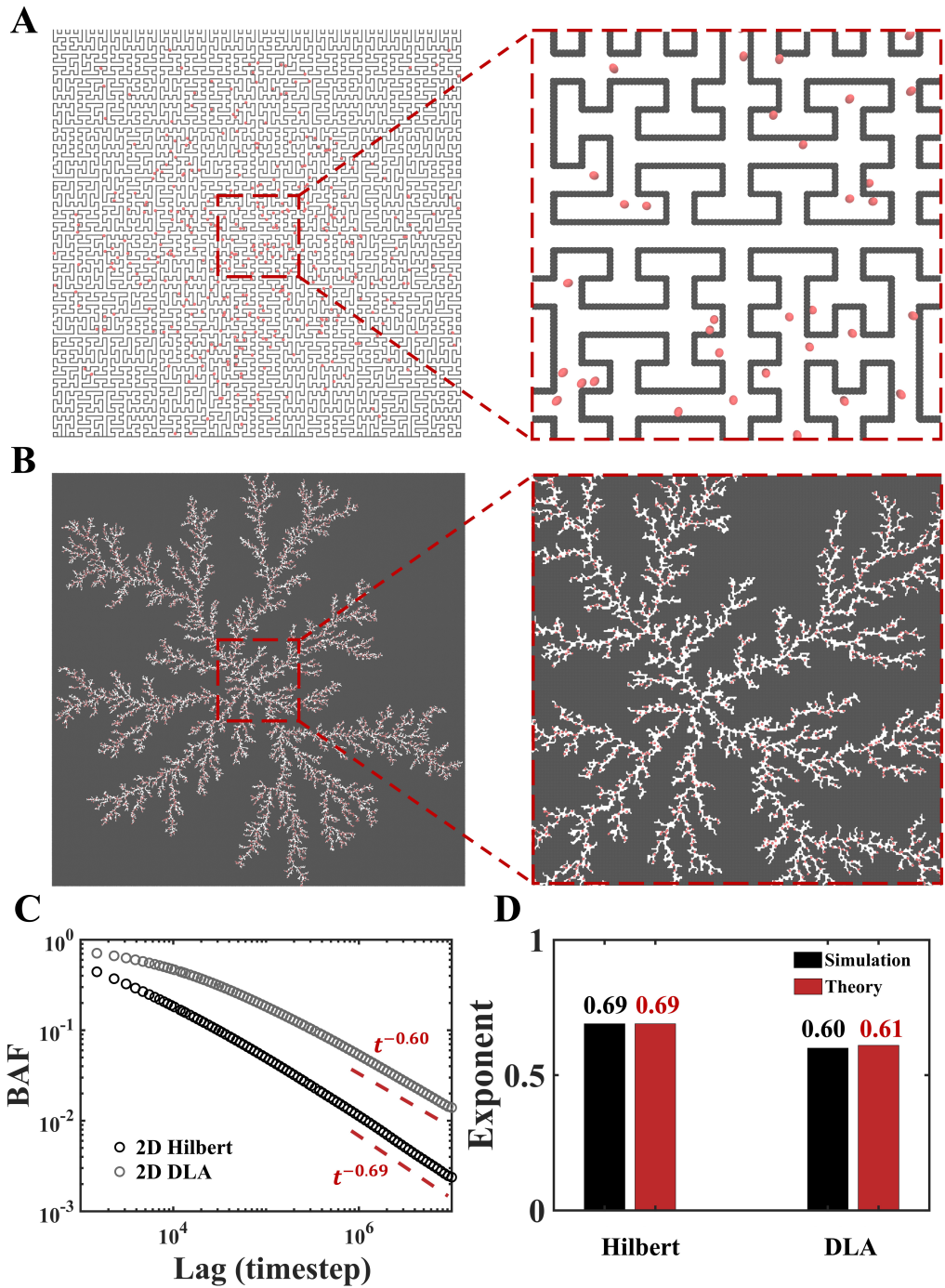

**Fig 3. Binding memory in fractal systems with anomalous diffusion.** (**A**) A liquid system of dynamic binding particles walled by a Hilbert curve. The particles are shown in red. Zooming in the system does not change the fractal structure. (**B**) A different fractal system constructed by two-dimensional diffusion-limited aggregation. (**C**) BAFs of the two fractal systems display power-law behaviors with different scaling exponents. (**D**) Agreement between simulation and theory on the scaling of BAFs in fractal systems.

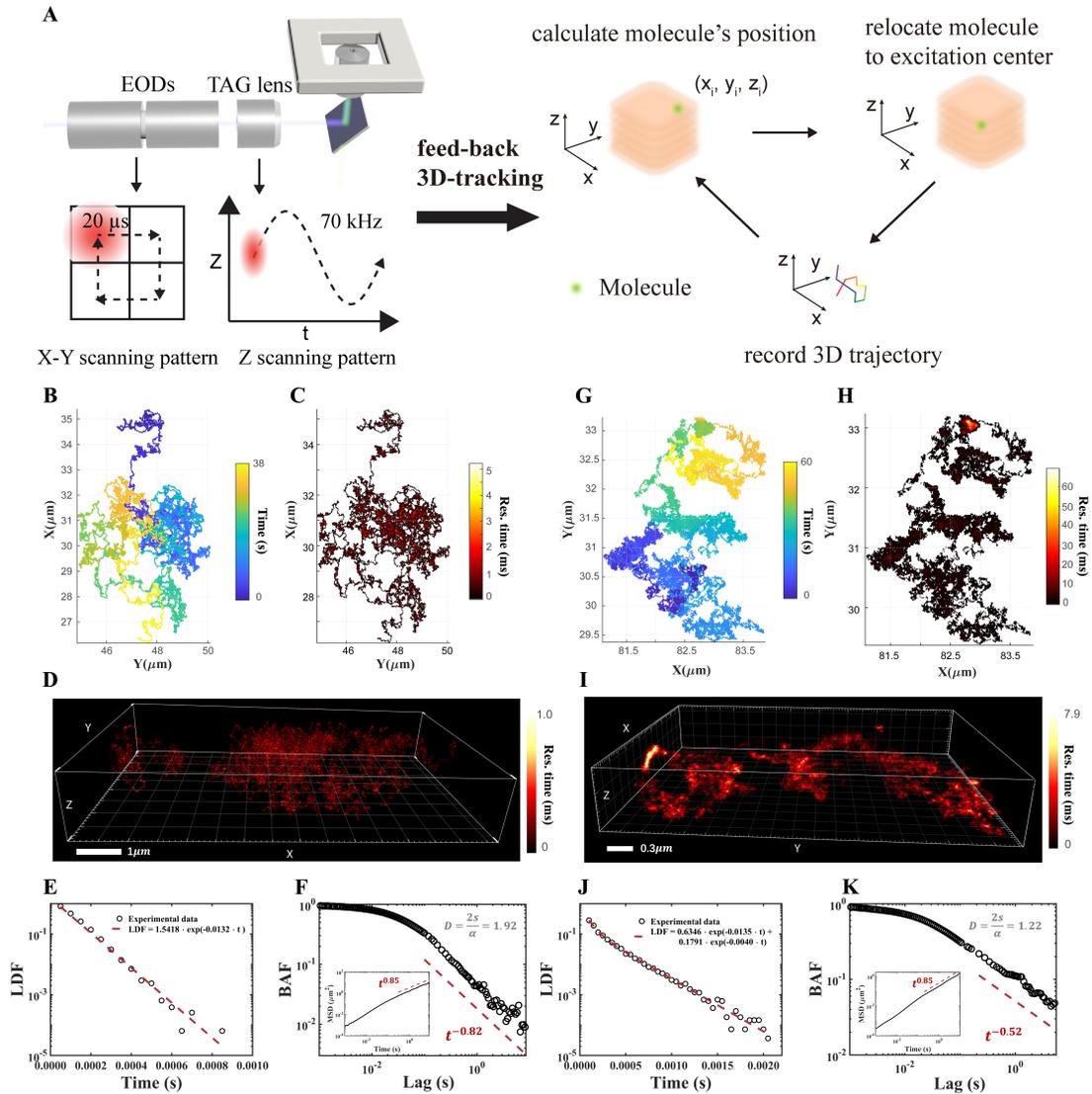

**Fig 4. Single-Particle Tracking experiments demonstrated binding memory effects on membranes.** (**A**) Schematic illustration of the 3D-SMARTER setup and the working procedure. (**B**) Trajectory of gp120-AgNP diffusing on a supported lipid bilayer. (**C, D**) 2D and 3D density maps of the residence time for the gp120-AgNP. (**E**) LDF of gp120-AgNP with a single-exponential fit. (**F**) BAF of gp120-AgNP exhibits a power law at long-time scale. The inset displays the mean squared displacement analysis, indicating a subdiffusive exponent, $\alpha = 0.85$. Based on Eq. 7, the fractal dimension of the lipid bilayer is inferred to be 1.92. (**G**) Trajectory of TATp-AgNP on a live-cell membrane. (**H, I**) 2D and 3D density maps of the residence time for TATp-AgNP. (**J**) LDF of TATp-AgNP on the live-cell membrane, with a double-exponential fit. (**K**) Power-law BAF and the MSD analysis inset, with the fractal dimension estimated to be 1.22 based on the theoretical model.

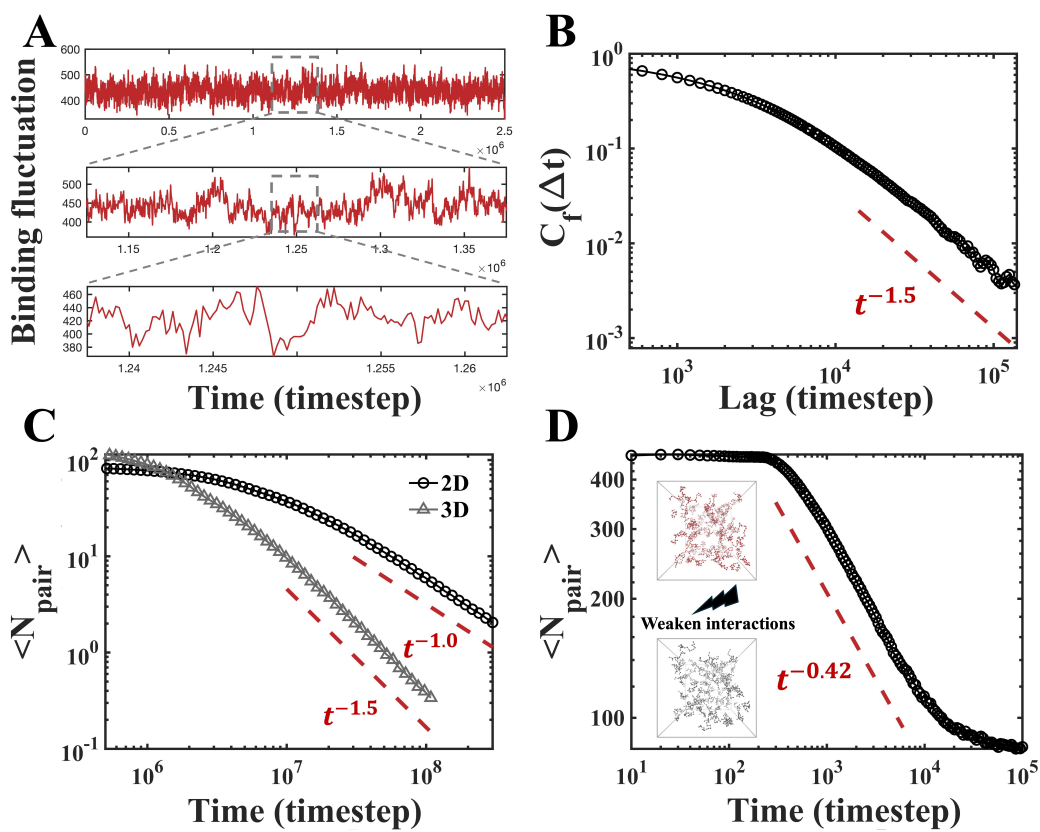

**Fig 5. Collective behaviors of binding memory** (**A**) Temporal fluctuations in the total number of bound molecules across different timescales. (**B**) Autocorrelation function analysis of binding fluctuations. (**C**) Time evolution of the total binding pairs between initially selected molecules in 2D and 3D systems, both showing power-law tails. (**D**) Response of the binding pairs to a sudden weakening of binding affinity in a nonequilibrium simulation, exhibiting a power-law behavior during the transition. In the inset, the high-affinity state of the molecules is shown in red and the low-affinity state in black.

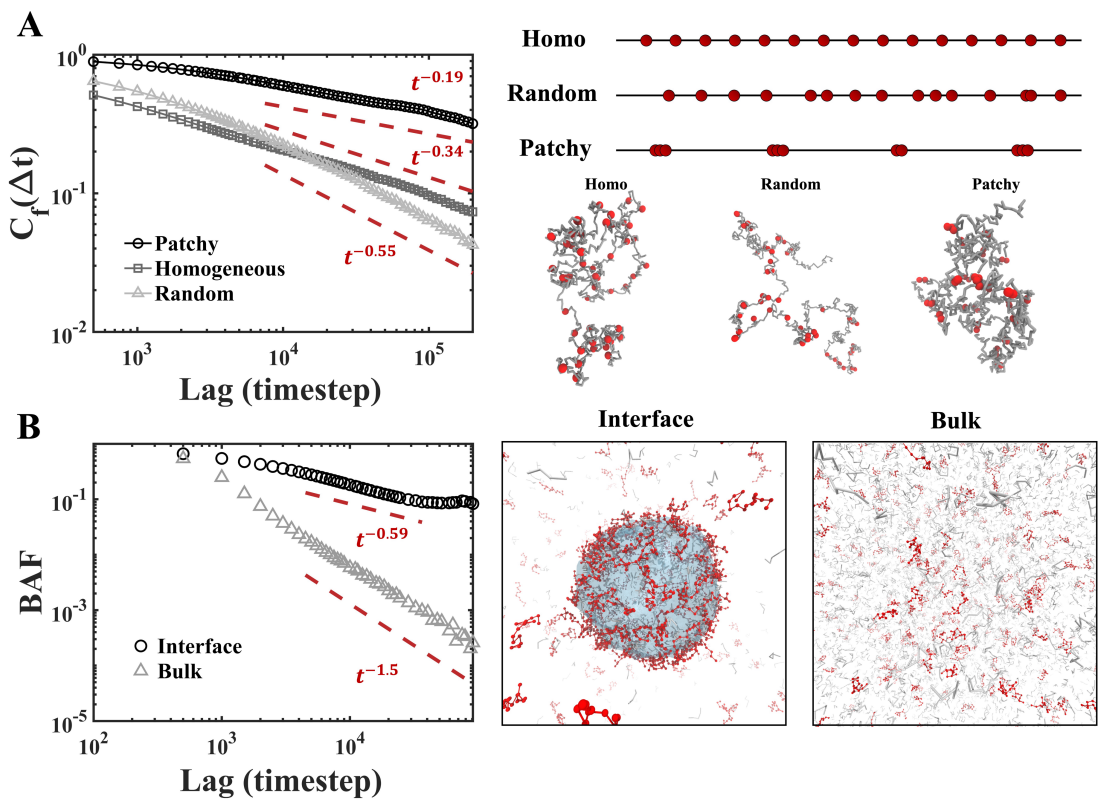

**Fig 6. Biological implications of molecular binding memory** (**A**) Fluctuation autocorrelation function of the number of binding pairs formed between stickers in a sticker-spacer polymer model with three different sequence types. While all sequences feature power-law behavior, they exhibit significantly different scaling exponents. The sequence patterns and corresponding typical conformations are shown on the right. (**B**) BAF of molecules adsorbed on the surface of a liquid droplets, in comparison with binding dynamics in bulk. Molecules at the droplet interface exhibit a stronger binding memory effect. Typical simulation snapshots of the two different systems are shown on the right.